\documentclass[useAMS,usegraphicx,usenatbib]{mn2e}
\usepackage{times}   %for Times Roman font (used by MNRAS)
\usepackage{mathptm} %Type1 fonts for mathematical symbols
\usepackage{color}
\usepackage{amsmath}

\title[Entropy-based dissipation trigger for SPH]
{A simple, entropy-based dissipation trigger for SPH}
\author[Rosswog ]
{S. Rosswog\thanks{E-mail: stephan.rosswog@astro.su.se}$^{1}$\\
$^1$ The Oskar Klein Centre, Department of Astronomy,  
Stockholm University, AlbaNova, SE-106 91 Stockholm, Sweden}

\def\be{\begin{equation}}
\def\ee{\end{equation}}
\def\bi{\begin{itemize}}

\def\ei{\end{itemize}}
\def\ben{\begin{enumerate}}
\def\een{\end{enumerate}}
\def\bea{\begin{eqnarray}}
\def\eea{\end{eqnarray}}
\def\bt{\begin{tabbing}}
\def\et{\end{tabbing}}

\def\edo{\end{document}}

\def\M4{M$_4$}
\def\M6{M$_6$}
\def\lcm6{LCM$_6$}
\def\qcm6{QCM$_6$}

\def\wh3{W$_{\rm H,3}$}
\def\wh4{W$_{\rm H,4}$}
\def\wh5{W$_{\rm H,5}$}
\def\wh6{W$_{\rm H,6}$}
\def\wh7{W$_{\rm H,7}$}

\newcommand{\Ma}{\texttt{MAGMA2}\,}
\newcommand{\ma}{\texttt{MAGMA2}}
\newcommand{\cd}{$d (\nabla \cdot \vec{v})/dt$}
\newcommand{\CD}{$d (\nabla \cdot \vec{v})/dt$ }

\begin{document}

\date{Draft version}

\pagerange{\pageref{firstpage}--\pageref{lastpage}} \pubyear{2013}

\maketitle

\label{firstpage}

\begin{abstract}
Smoothed Particle Hydrodynamics (SPH) schemes need to be enhanced by
dissipation mechanisms to handle shocks. 
Most SPH formulations rely on artificial viscosity and while this 
working well in pure shocks, attention has to be payed to avoid dissipation
where it is not wanted. Commonly used approaches include limiters 
and time-dependent dissipation parameters. The former try to distinguish
shocks from other types of flows that do not require dissipation while in the
latter approach the dissipation parameters are steered by some source term ("trigger") 
and, if not triggered, they decay to a pre-described floor value.
The commonly used source terms  trigger on either compression, $-\nabla\cdot\vec{v}$, or its 
time derive.  Here we explore a novel way to trigger SPH-dissipation: 
since an ideal fluid conserves entropy exactly, its numerical non-conservation
can be used to identify "troubled particles" that need dissipation because
they either pass through a shock or become noisy for other reasons.
Our new scheme is implemented into the Lagrangian hydrodynamics code \Ma 
and is scrutinized in a number of shock and fluid instability tests. We find 
excellent results in shocks and only a very moderate (and desired) switch-on 
in instability tests. The new scheme is robust, trivial to implement into existing 
SPH codes and does  not add any  computational overhead.
\end{abstract}

\section{Introduction}
Smoothed Particle Hydrodynamics \citep{lucy77,monaghan77} is a completely mesh-free 
method to solve the equations of hydrodynamics. It can be elegantly derived from a discretized
Lagrangian of an ideal fluid \citep{gingold82,speith98,monaghan01,springel02,rosswog15c} and thus ensures that
Nature's conservation laws are obeyed. As derived in this way, SPH is entirely dissipationless, 
and can therefore not handle shocks: in a shock front bulk kinetic energy is transformed 
by dissipation into internal energy which goes along with an increase in entropy. 
Therefore, the ideal SPH equations need to be enhanced by some dissipative mechanism.
 In most modern Eulerian hydrodynamics schemes this is achieved by applying (exact or 
 approximate) Riemann solvers, see e.g. \cite{toro99}. This is also possible in SPH 
\citep{inutsuka02,cha03,puri14}, but the use of artificial viscosity is more widespread. It is
worth mentioning, however, that many artificial viscosity schemes bear similarities with approximate
Riemann solvers \citep{monaghan97}. While Riemann solvers are an elegant concept and less 
ad hoc than artificial viscosity, 
they always provide some amount of dissipation even in situations where it would actually not 
be needed. In SPH one has (at least in principle) the possibility to suppress/switch off dissipation
completely. Historically, however, early implementations applied artificial viscosity terms
with constant parameters and without limiters so that dissipation was always switched on whether
needed or not.This lead to excessively dissipative SPH schemes and controlling
the amount of dissipation has been a concern since.\\
Suggested cures include "limiters" \citep{balsara95,cullen10,read12,wadsley17} that are aimed at suppressing
dissipation outside of shocks, tensor prescriptions  \citep{owen04} that intend
to restrict the effects of artificial viscosity to the shock travel directions and dissipation schemes
with time dependent parameters. Time dependent dissipation
parameters were introduced by \cite{morris97} who suggested to evolve them separately for 
each particle according to an additional differential equation with a source and a decay term 
that, unless triggered, drive the parameter to a pre-defined small floor value,
see Eq.~(\ref{eq:davdt}) below.
As a source term they used $-\nabla\cdot \vec{v}$ which works well in many cases, but cannot distinguish
between an adiabatic compression and an entropy producing shock. \cite{cullen10} suggested to use  instead 
$d(-\nabla\cdot \vec{v})/dt$ as a dissipation trigger, so that a particle that moves into a shock (and thereby becomes increasingly 
more compressed) raises its dissipation parameter which subsequently decays once the shock wave has passed.\\
Here we explore an alternative trigger that involves keeping track  of some entropy measure at the particle
level. Monitoring entropy violations has been used to steer dissipation in
Eulerian Newtonian hydrodynamics \citep{guermond11,guermond16} and it has also been used in relativistic
hydrodynamics to steer to which amount low-order fluxes need to be added to higher-order fluxes
for numerical stabilization \citep{guercilena17}. In this paper we use the local violation of exact 
entropy conservation to steer how much dissipation every SPH-particle needs. As shown in the tests
below, our scheme yields excellent results, is trivial to implement and comes without any computational
overhead. We describe our methodology in Sec.~\ref{sec:meth}, where we also briefly summarize the
key ingredients of our \Ma code  in Sec.~\ref{sec:SPH}, and we discuss 
the entropy dissipation trigger in Sec.~\ref{sec:steering}. In Sec.~\ref{sec:tests} we show a number 
of benchmark tests and we conclude in Sec.~\ref{sec:summary}  with
a concise summary.

\section{Methodology}
\label{sec:meth}
\subsection{The Smoothed Particle Hydrodynamics formulation}
\label{sec:SPH}
The SPH code \Ma  \citep{rosswog20a} profits from a number of new elements: a) it uses of high order
kernels, b) calculates accurate gradients via matrix inversion
techniques and c) uses a new dissipation scheme where velocities are reconstructed
via slope limiter techniques to the inter-particle midpoint \citep{christensen90,frontiere17}.
The differences of these
reconstructed velocities are used in the artificial viscosity tensor rather than the ("flat") differences
of the particle velocities, as is the standard practice in SPH. This approach drastically
reduces unwanted dissipation and we have shown in an extensive set of test cases \citep{rosswog20a}
that excellent results are obtained even if the dissipation parameter $\alpha$ is kept
constant at its maximum value. This finding is consistent with the results of \cite{frontiere17} who 
used an SPH formulation based on  reproducing kernel interpolation \citep{liu95}.
The main aim of this study is to apply dissipation only to 
"troubled particles" that are identified via entropy non-conservation between two subsequent
time steps.\\
The equation set that we are using has been developed and tested extensively in a special
relativistic context \citep{rosswog15b} and --in its Newtonian version-- in the recent \Ma code 
paper \citep{rosswog20a}.
We use
\bea
\rho_a&=& \sum_b m_b W_{ab}(h_a),
\label{eq:dens_sum}\\
\frac{d\vec{v}_a}{dt}&=& - \sum_b m_b \left\{ \frac{P_a}{\rho_a^2} \vec{G}_a +  \frac{P_b}{\rho_b^2} \vec{G}_b\right\}
\label{eq:momentum_IA},\\
\left(\frac{du_a}{dt}\right)&=& \frac{P_a}{\rho_a^2} \sum_b m_b \vec{v}_{ab} \cdot \vec{G}_a
 \label{eq:energy_IA},
\eea
as density, momentum and energy equation,
where $\rho, \vec{v}_a, u_a$ denote mass density, velocity and specific internal energy, $m$ is the particle mass, $P$
the gas pressure, $\vec{v}_{ab}= \vec{v}_a - \vec{v}_b$, $W$ the chosen SPH kernel function and $h$
its smoothing length.
The gradient functions are given by
\be
\left(\vec{G}_{a}\right)^k= \sum_{d=1}^D C^{kd}(\vec{r}_a,h_a) (\vec{r}_b - \vec{r}_a)^d W_{ab}(h_a),
\ee
\be
\left(\vec{G}_{b}\right)^k= \sum_{d=1}^D C^{kd}(\vec{r}_b,h_b) (\vec{r}_b - \vec{r}_a)^d W_{ab}(h_b),
\ee
where $C$ is a "correction matrix" that accounts for the local particle distribution and is calculated as
\be
\left(C^{kd}_a (h)\right)= \left( \sum_b \frac{m_b}{\rho_b} (\vec{r}_b - \vec{r}_a)^k (\vec{r}_b - \vec{r}_a)^d W(|\vec{r}_a-\vec{r}_b|,h)\right)^{-1}.
\label{eq:corr_mat}
\ee
Such gradients have been shown to work well \citep{garcia_senz12,cabezon12a} and to be orders of magnitude more
accurate than standard SPH-kernel gradient methods, see Fig. 1 in \cite{rosswog15b}.
Following the approach of \cite{vonNeumann50}, we implement artificial viscosity 
by adding an "artificial pressure" $Q$ to the physical pressure  $P$ wherever it occurs.
We use the expression \citep{monaghan83} 
\be
Q_a= \alpha \; \rho_a \left(-  c_{{\rm s},a} \mu_a +  2 \mu_a^2\right),
\label{eq:Qvis}
\ee
where the velocity jump is (summation over $\delta$)
\be
\mu_a= {\rm min} \left(0, \frac{v_{ab}^\delta \eta_a^\delta }{\eta_a^2 + \epsilon^2}\right).
\label{eq:mu_vis}
\ee
The numerical parameters $\alpha$ and $\epsilon$ are usually set to 1 and $0.1$,
$c_{{\rm s},a}$ is the sound speed and 
\be
\eta^\delta_a= \frac{\left(\vec{r}_a - \vec{r}_b\right)^\delta}{h_a}, \quad \eta_a^2=  \eta^\delta_a  \eta^\delta_a
\ee
are (non-dimensionalized) separations between particles.
In SPH it is common practice to use $v_{ab}^\delta = \vec{v}_a^\delta -  \vec{v}_a^\delta$ in Eq.~(\ref{eq:mu_vis}), 
i.e. to apply the velocity difference between the two particles.  In \Ma we quadratically reconstruct the velocities
of particle $a$ and $b$ to their midpoint at $\vec{r}_{ab}= 0.5(\vec{r}_a+\vec{r}_b)$, so that the velocities reconstructed from
the  $a$-side read
\be
\tilde{v}_a^i = v_a^i + \Phi_{ab} \left[ (\partial_j v^i) \delta^j  + \frac{1}{2} (\partial_l \partial_m v^i) \delta^l \delta^m \right]_a,
\ee
where the index at the square bracket indicates that the derivatives
at the position of particle $a$ are used and the increments
from point $a$ to the midpoint are $(\delta^i)_a= \frac{1}{2} (\vec{r}_b - \vec{r}_a)$. The reconstructed 
velocities from the $b$-side, $\tilde{v}_b^i$, are calculated correspondingly,
but with derivatives from position $b$ and increments $\delta^i_b= -\delta^i_a$. In Eq.~(\ref{eq:mu_vis}) we use the
difference in the {\em reconstructed} velocities, i.e. 
 $v_{ab}^\delta= \tilde{v}_a^\delta - \tilde{v}_b^\delta$. 
 To calculate the first and second velocity derivatives we also use matrix inversion techniques,  see
\cite{rosswog20a} for more details.
\\
%-----------------------------------------------------------------------                                                                
\begin{figure}
   \centering
   \centerline{\includegraphics[width=\columnwidth,angle=0]{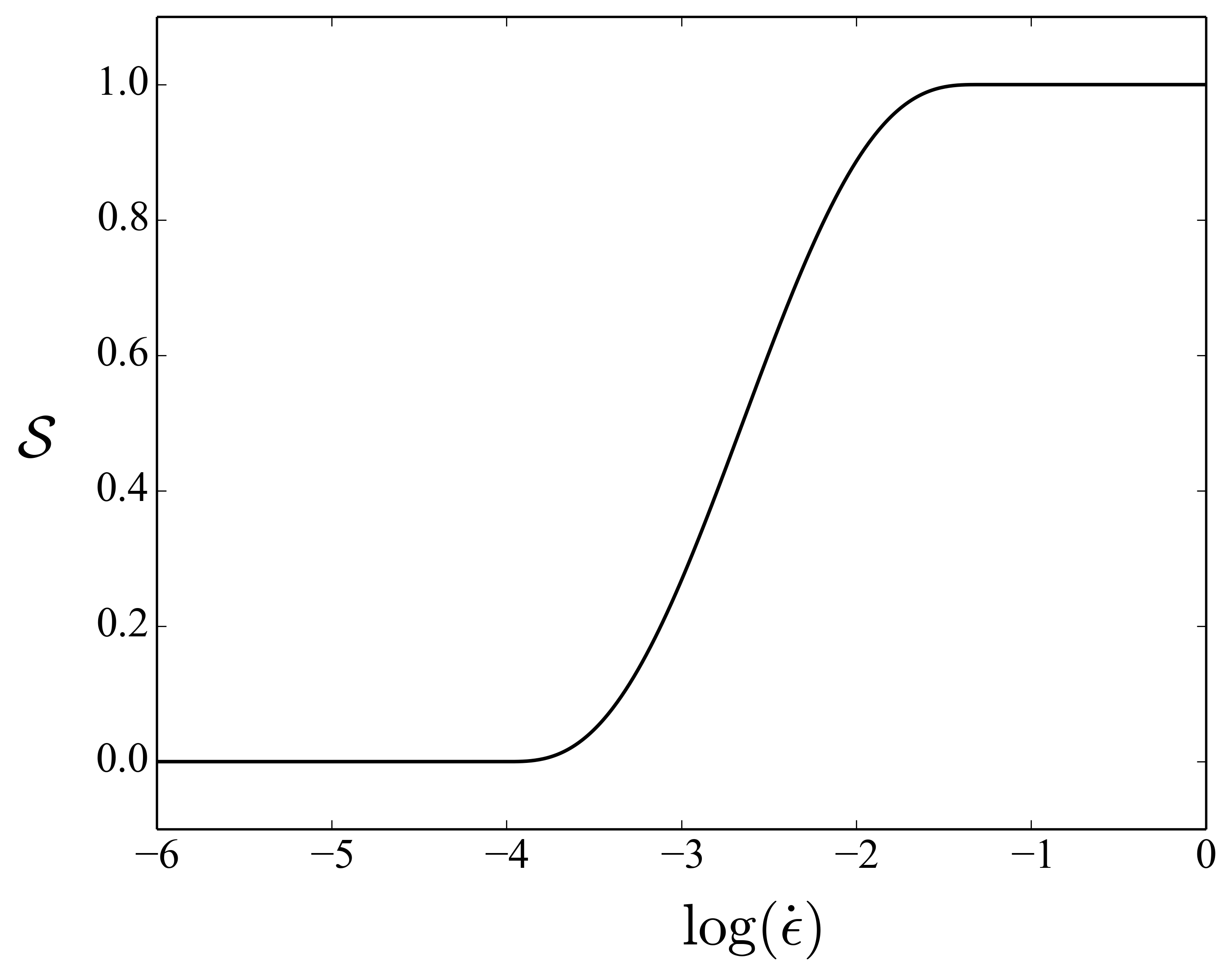}}
    \caption{The smooth switch-on function that we use to translate entropy violations into dissipation parameter values, see Eq.~(\ref{eq:switch_on}).}
   \label{fig:switchon}
\end{figure}
%-----------------------------------------------------------------------      
%-----------------------------------------------------------------------------
\begin{figure*}
\hspace*{-0.cm}\includegraphics[width=2.1\columnwidth,angle=0]{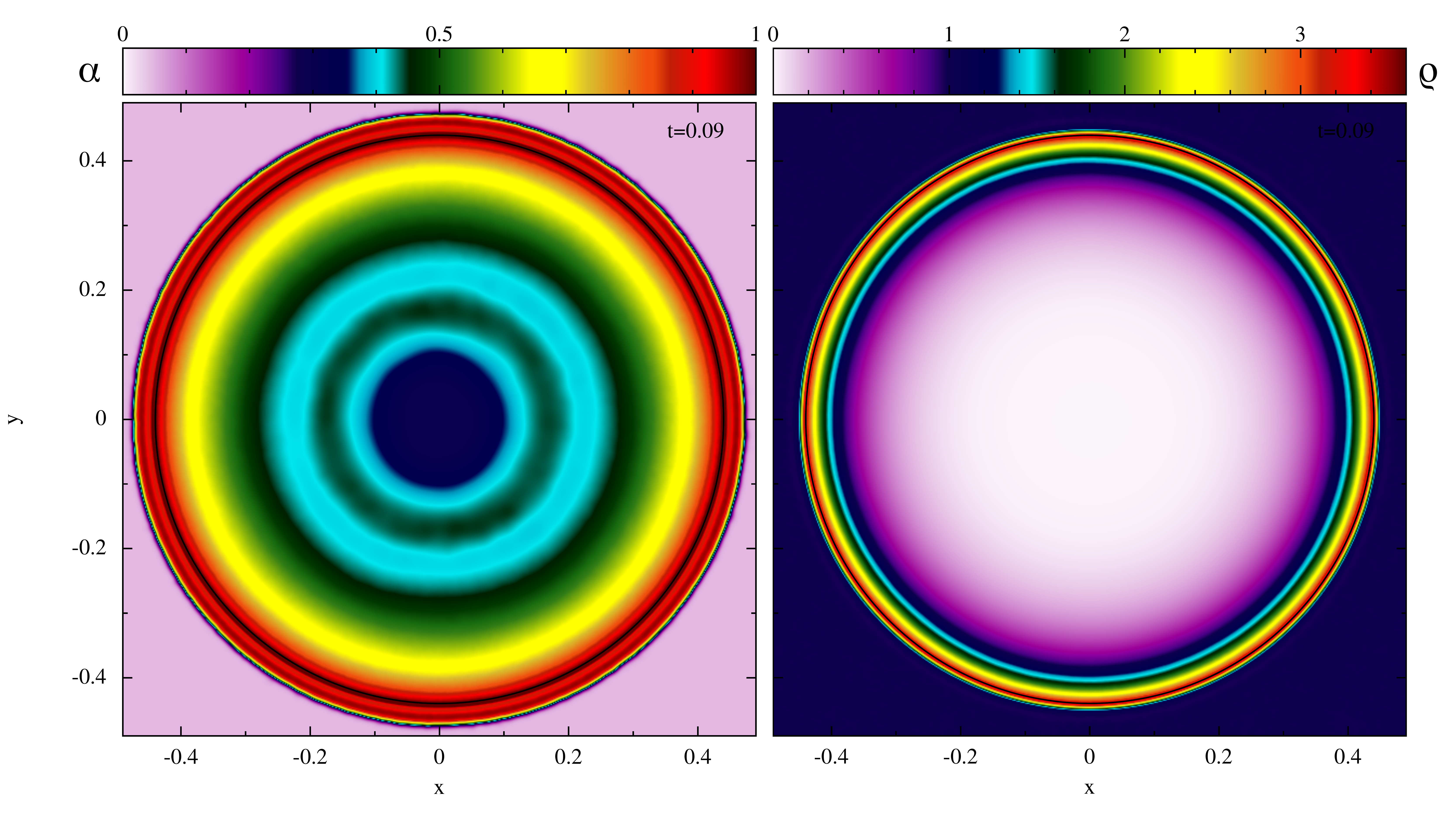}
\vspace*{0cm}
\caption{3D Sedov test $(200^3$ particles), dissipation parameter $\alpha$ on the left and density $\rho$ on the right at $t=0.09$.}
\label{fig:Sedov1}
\end{figure*}
%-------------------------------------------------------------------------------
%-----------------------------------------------------------------------------
\begin{figure*}
\hspace*{0.cm}\includegraphics[width=2.\columnwidth,angle=0]{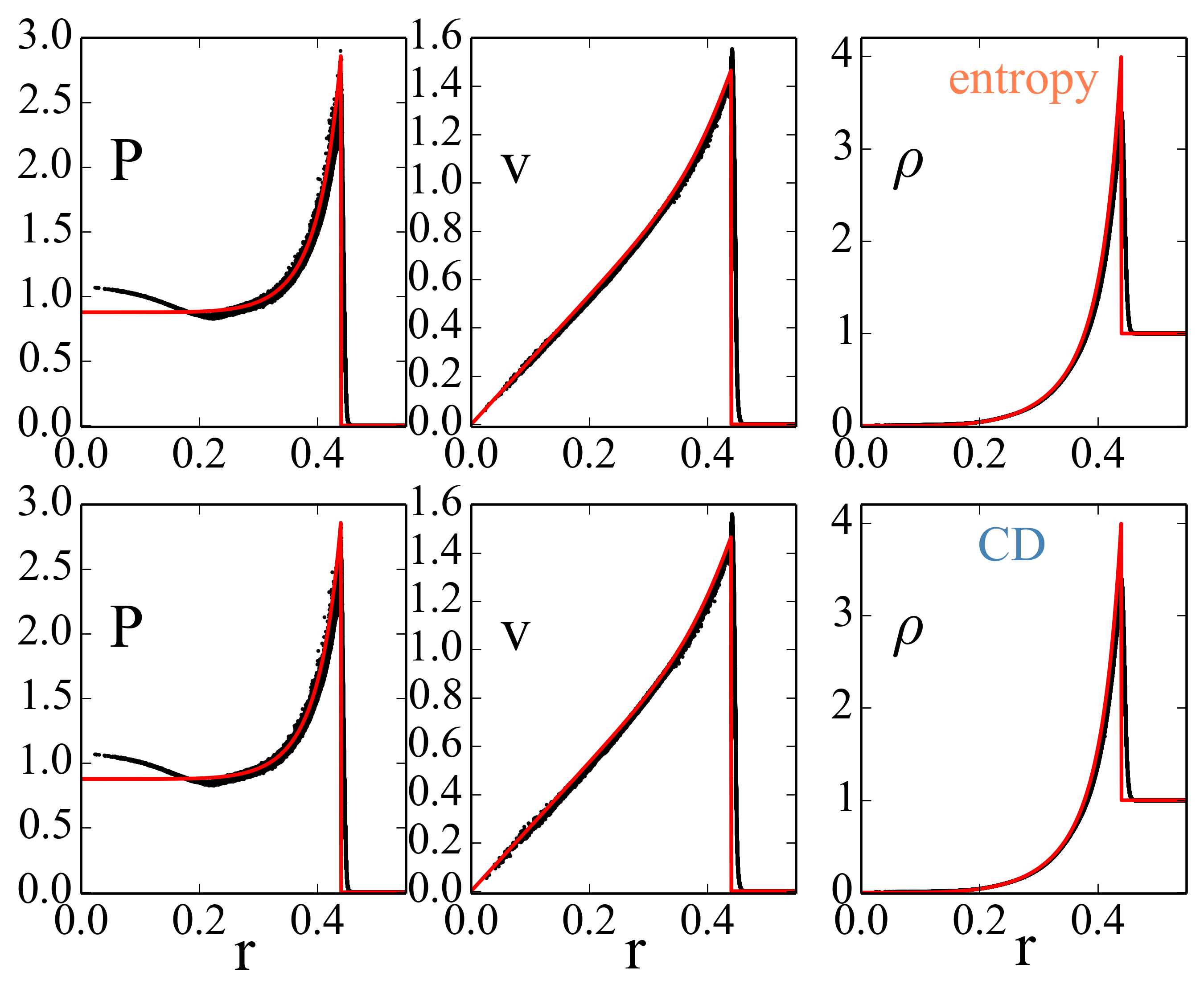}
\caption{3D Sedov test $(200^3$ particles, at $t=0.09$); upper row: the suggested entropy-steering; lower row: \CD-steering. Each time
pressure $P$  (left), velocity $v$ (middle) and  density $\rho$(right), all particles (black) and the exact solution (red) are shown.}
\label{fig:Sedov2}
\end{figure*}
%-------------------------------------------------------------------------------
%-----------------------------------------------------------------------------
\begin{figure}
\hspace*{0.cm}\includegraphics[width=1\columnwidth,angle=0]{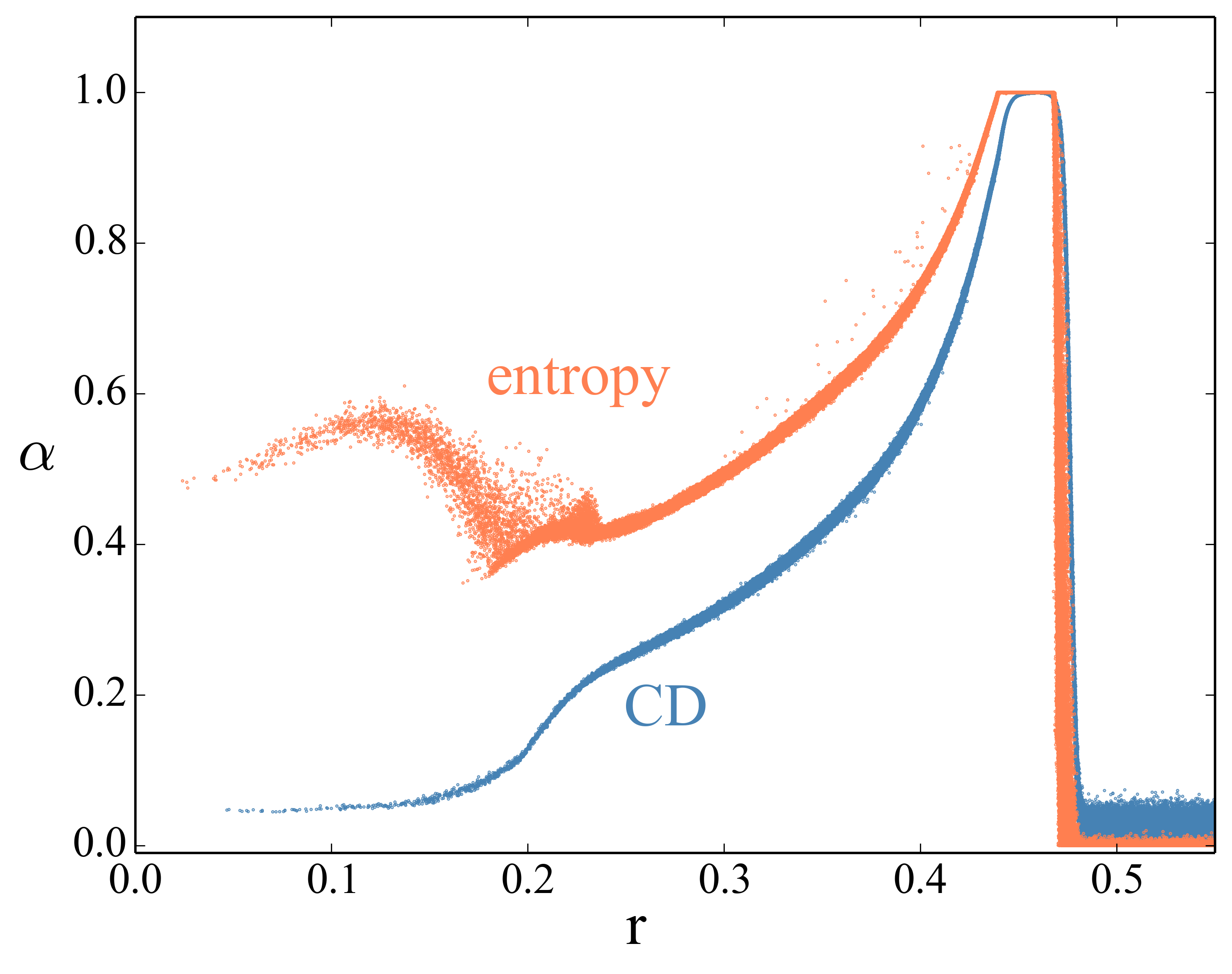}
\caption{Comparison of the triggered dissipation parameter values in 3D Sedov test $(200^3$ particles) between the proposed scheme ("entropy") and the \CD  trigger (CD).}
\label{fig:Sedov3}
\end{figure}
%-------------------------------------------------------------------------------
%-----------------------------------------------------------------------------
\begin{figure*}
\includegraphics[width=2.\columnwidth,angle=0]{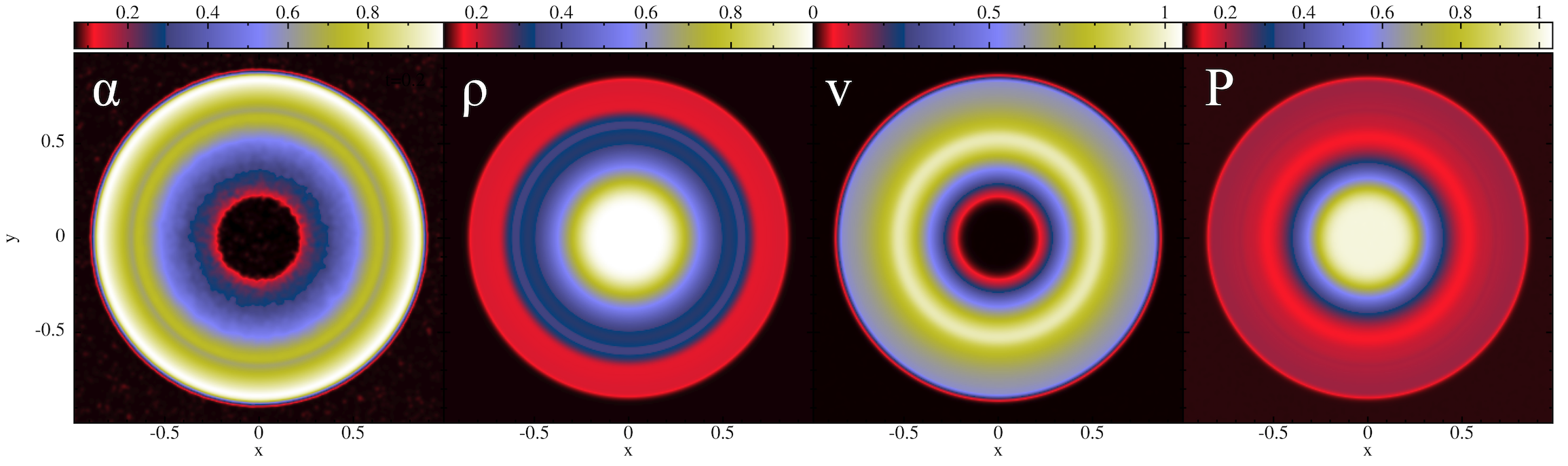}
\caption{Spherical Riemann problem in 3D with entropy trigger. From left to right: dissipation parameter 
$\alpha$, density $\rho$, velocity $v$ and pressure $P$.}
\label{fig:Riemann1}
\end{figure*}
%-------------------------------------------------------------------------------
%
%-----------------------------------------------------------------------------
\begin{figure*}
\includegraphics[width=2\columnwidth,angle=0]{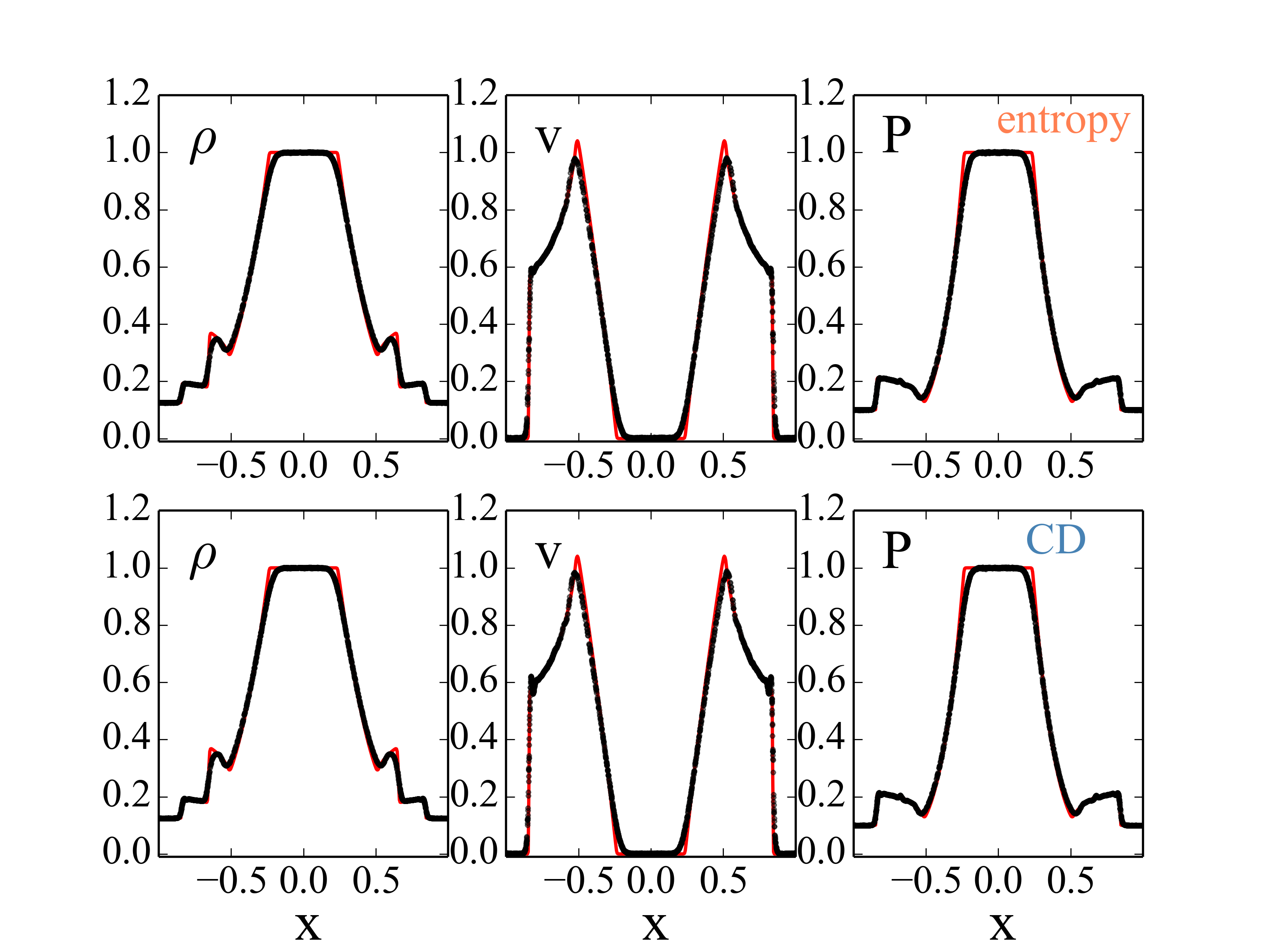}
\caption{Spherical Riemann problem in 3D: the \Ma solution ($200^3$ particles) is 
shown as black circles, the reference solution (red line) has been obtained with  the 
Eulerian weighted average flux method with $400^3$ grid cells  (Toro 1999). The upper
row shows the results from the entropy trigger, the lower one is obtained using the \CD-trigger.}
\label{fig:Riemann1_ref}
\end{figure*}
%-------------------------------------------------------------------------------
We use a  modification of van Leer's slope limiter \citep{vanLeer74,frontiere17}
\be
\Phi_{ab}= \rm{max}\left[0, \rm{min}\left[ 1, \frac{4A_{ab}}{(1+A_{ab})^2}\right] \right] \begin{cases}
    1, & \text{if $\eta_{ab} > \eta_{\rm{crit}}$}.\\
    e^{-\left(\frac{\eta_{ab}-\eta_{\rm crit}}{0.2}\right)^2}, & \text{otherwise}
  \end{cases}
\ee
with
\be
A_{ab}= \frac{(\partial_\delta v_a^\gamma) \; x_{ab}^\delta \; x_{ab}^\gamma}{(\partial_\delta v_b^\gamma) \; x_{ab}^\delta \; x_{ab}^\gamma}
\ee  
and
\be
\eta_{ab}= \rm{min}(\eta_a,\eta_b) =  \rm{min}\left(\frac{r_{ab}}{h_a},\frac{r_{ab}}{h_b}\right)  \; {\rm and} \;
\eta_{\rm crit}= \left( \frac{32\pi}{3N_{\rm nei}}\right)^{1/3},
\ee
with $N_{\rm nei}$ being the number of neighbours for the chosen kernel. We also apply  a small 
amount of thermal conductivity
\be
\left( \frac{du_a}{dt}\right)_C= -\alpha_u \sum_b m_b \frac{v_{\rm sig,u}^{ab}}{\rho_{ab}} \left(u_a - u_b \right)
\frac{|\vec{G}_a+\vec{G}_b|}{2},
\ee
where $\alpha_u= 0.05$ and 
$v_{\rm sig,u}$ is a signal velocity. For more details and  the explicit expressions that we use we refer to the 
 \Ma code paper \citep{rosswog20a}. In all of the shown tests we use the Wendland C6 kernel \citep{wendland95} 
 with 300 neighbour particles.

\subsection{Using entropy non-conservation to identify "troubled particles"}
\label{sec:steering}
Our SPH formulation conserves mass energy, momentum and 
angular momentum exactly\footnote{Modulo effects from the grad-h effects, potential finite accuracy in the 
ODE integration and approximations for gravitational forces etc.}, entropy conservation, in contrast, 
is not actively enforced and therefore its potential non-conservation can be used to monitor the smoothness
of the local flow. In smooth flows entropy should be conserved exactly while it may be physically increased 
in shocks. Flows can, however, also become "noisy" (i.e. non-negligle velocity fluctuations appear)
for numerical reasons (e.g. particles moving off an initially specified, non-optimal lattice) and also
in such cases (a smaller amount of) dissipation is desirable.
In either case, shocks or noisy flows, 
one would want to apply artificial dissipation in order to keep the flow physically well-behaved, and 
measuring the degree of numerical non-conservation of entropy
(or some entropy function) is a natural way  to identify "troubled particles" and  to determine 
how much dissipation should be applied.\\ 
Here we suggest to measure the {\em rate} of numerical entropy generation
between two subsequent time steps and a translation of this rate into
a value for the dissipation parameter $\alpha$. Since \Ma produces, due to the velocity reconstruction, 
excellent results even with a constant $\alpha=1$, we choose our parameters conservatively large
so that $\alpha$ reaches already substantial values for small entropy violations. For SPH schemes
without such a velocity reconstructions the same functional relations can be used, but the optimal
parameter values may have slightly different values.\\   
\noindent We assume here a polytropic equation of state and use 
\be
s_a\equiv \frac{P_a}{\rho_a^\Gamma} 
\ee
as a measure for the entropy carried by particle $a$. Here $P_a$ is the gas pressure and $\Gamma$ 
the polytropic exponent. Polytropic equations of state are used in most astrophysical
gas simulations, but other entropy measures, e.g. the physical entropy of an ideal gas, could 
equally well be used along the same lines of reasoning. Even if other sources of entropy 
(e.g. nuclear reactions) should be present, this approach can be used provided that one
can cleanly separate out the contributions from the additional source. But this is not the topic
of our study here and we leave this for future investigations.\\
We use the non-dimensionalized relative 
entropy rate of change of a particle $a$ between time step $t^{n-1}$ and time step $t^{n}$ ($\Delta t= t^{n} - t^{n-1}$)
\be
\dot{\epsilon}_a^n \equiv \frac{ |s_a^{n} - s_a^{n-1}|}{s_a^{n-1}} \frac{\tau_a}{\Delta t},
\ee
as a measure of how much dissipation is needed. Here
$\tau_a= h_a/c_{s,a}$ is the particle's dynamical time scale and $c_{s,a}$ its sound speed.
We use  $l_a^n\equiv \log(\dot{\epsilon}_a^n)$ to steer the amount of dissipation.
Similar to earlier work \citep{morris97,rosswog00,cullen10,rosswog15b,wadsley17},
we  let the dissipation parameter  $\alpha_a$ decay according to
\be
\frac{d\alpha_a}{dt}=  -\frac{\alpha(t) - \alpha_0}{30 \tau_a},
\label{eq:davdt}
\ee
where $\alpha_0$ is a floor value, in other schemes often set to values
around 0.1 to keep the particle distribution well-behaved \citep{tricco19a}. 
Note that we have conservatively chosen a rather long decay time scale of 30 $\tau_a$.
We compare at each time step the actual value to  a "desirable dissipation parameter"  
and if the latter exceeds the current value, $\alpha(t)$ is increased instantly \citep{cullen10}.
The desired value  of $\alpha$ is chosen according to  the trigger $l_a^n$
\be
\alpha_{a,\rm des}^n= \alpha_{\rm max} \; \mathcal{S}(l_a^n),
\ee
where $\mathcal{S}$ is smooth "switch-on" function for which we have
chosen, see Fig.~\ref{fig:switchon},
\be
\mathcal{S}(x)= 6x^5 - 15x^4 + 10x^3,
\label{eq:switch_on}
\ee
with 
\be
x= \rm min\left[max\left(\frac{l_a^n - l_0}{l_1-l_0},0\right),1\right].
\ee
The reasoning behind this is that we consider dissipation unnecessary for acceptably small
entropy violations (exact value set by $l_0$) and beyond another threshold value (set by $l_1$)
we need the maximal dissipation parameter, $\alpha_{\rm max}$.
After some experimenting with both shocks and instability tests, we  settled on values 
$l_0= \log(1 \times 10^{-4})$ and $l_1= \log(5 \times 10^{-2})$, so that our scheme
does not switch on at all for entropy violations $\dot{\epsilon}_a^n \le 10^{-4}$ and reaches
$\alpha= \alpha_{\rm max}= 1$ for $\dot{\epsilon}_a^n \ge 5 \times 10^{-2}$, see Fig.~\ref{fig:switchon}.
For aesthetic reasons we prefer to have only {\em triggered} dissipation rather than assigning
a floor value $\alpha_0$ by hand. We therefore use $\alpha_0=0$ in our implementation, but note that with the
chosen parameters $l_0$ and $l_1$ a small amount of dissipation (typically $\alpha \sim 0.01$) is triggered
even in smooth flows, see below. We find good results for this particular switch-on function 
and the chosen parameters, but other choices are certainly possible and the optimal parameter values
might be slightly different for other SPH formulations.\\
In the below tests, we compare also to the \cd-trigger suggested by \cite{cullen10}.
Here the desired dissipation parameter is chosen as\footnote{Note that their kernels have a support radius of
1$h$ while we follow the convention that the kernel is non-zero out to a radius of 2$h$.}
\be
\alpha_{a,\rm des}^{n, \rm CD}= \frac{A_a}{0.25\left(\frac{v_{a, \rm sig}}{h_a}\right)^2 + A_a},
\ee
where $A$= min(-d$(\nabla\cdot\vec{v})$/dt,0) and the signal velocity is given by
\be
v_{a, \rm sig}= \rm{max}_b[c_{{\rm s},ab} - \rm{min}\{0,\vec{v}_{ab} \cdot \hat{e}_{ab}\}].
\ee
Here, $c_{{\rm s},ab}= 0.5(c_{{\rm s},a} + c_{{\rm s},b})$, $\vec{v}_{ab}= \vec{v}_a - \vec{v}_b$ and
$\hat{e}_{ab}= (\vec{r}_a-\vec{r}_b)/|\vec{r}_a-\vec{r}_b|$. \cite{cullen10} chose a decay 
time scale of $20 h_a/v_{a, \rm sig}$ for the denominator of Eq.~(\ref{eq:davdt}) and this is 
the parameter we adopt in this comparison. It is worth pointing out that this is
a comparison of {\em triggers} and there are differences between our approach with
the \CD-trigger (e.g. SPH-formulation, kernels, reconstruction, conductivity) and the 
original Cullen \& Dehnen approach. Nevertheless, we use the  "CD" 
in some of the below figures as a shorthand for this \CD-trigger approach.

\section{Tests}
\label{sec:tests}
To scrutinize the suggested scheme, we perform a number of benchmark tests.
We perform shock tests to demonstrate that the dissipation robustly switches on and avoids
spurious oscillations and Kelvin-Helmholtz and  Rayleigh-Taylor tests to verify that no unnecessary 
dissipation is triggered in smooth portions of the flow.
\subsection{Sedov Taylor Blast}
We begin by setting up a Sedov explosion test where a given number 
of SPH particles is distributed according to a centroidal Voronoi tesselation 
\citep{du99} in the computational volume $[-0.5,0.5]\times[-0.5,0.5]\times [-0.5,0.5]$.
While this already produces very good quality initial conditions, they can be further 
improved by applying regularization sweeps, where each particle position is corrected
according to Eq.~(20) of \cite{gaburov11}. This procedure ensures nearly perfectly spherically
symmetric results in this test. Subsequently we assign
masses so that the density is $\rho=1$.  We use a polytropic exponent
$\Gamma=5/3$ and spread an internal energy $E= 1$ across
a very small initial radius $R$, the
specific internal energy $u$ of the particles outside of $R$ is entirely negligible
($10^{-10}$ of the central $u$). For the initial radius $R$ we choose twice
the interaction radius of the innermost SPH particle. 
Boundaries play no role in this test (as long as the blast doesn't interact with them),
we therefore place "frozen" particles with fixed properties
around the computational volume as boundary particles. 
For more details we refer to \cite{rosswog20a}.\\
 We show in Fig.~\ref{fig:Sedov1} the dissipation parameter $\alpha$ (left) and 
the density $\rho$ (right) for  $200^3$ SPH particles
 (excluding boundary particles). This test
requires large dissipation values, both to robustly handle the shock and to "calm" 
the particles in the post-shock region. Our scheme delivers large $\alpha$-values
in this test with values of $\approx$ 1 in the shock itself, and a moderate decay to values
around 0.4 in the shocked, inner region. We show in Fig.~\ref{fig:Sedov2}, upper row, the numerical
solution of pressure, velocity and density (black dots) as a function of radius together with the exact solution
(red). Note that {\em all} particles are plotted.  Keep in mind that this numerical test 
challenges most methods whether particle- or mesh-based and often noise and strong spurious
post-shock oscillations occur, see  for example
\cite{rosswog07c,hu14a,cardall14,hopkins15a,wadsley17} or \cite{frontiere17}. 
For our results, the overall agreement is very good and 
there is only a small velocity overshoot at the shock and --since the finite 
particle masses-- overestimate the close-to-zero central density, the central pressure
is somewhat overestimated. \\
The lower row of Fig.~\ref{fig:Sedov2} shows the corresponding result for the \cd-trigger.
In this test the \CD-trigger delivers results that are very similar to the entropy trigger. It
provides somewhat less dissipation in the shocked inner region, see Fig.~\ref{fig:Sedov3},
and therefore the velocity distribution there is slightly more noisy.

%-----------------------------------------------------------------------------
%%%% in original labelling this is **Test 3** %%%%%%%%%%%%
\begin{figure*}
\centerline{\includegraphics[width=2\columnwidth,angle=0]{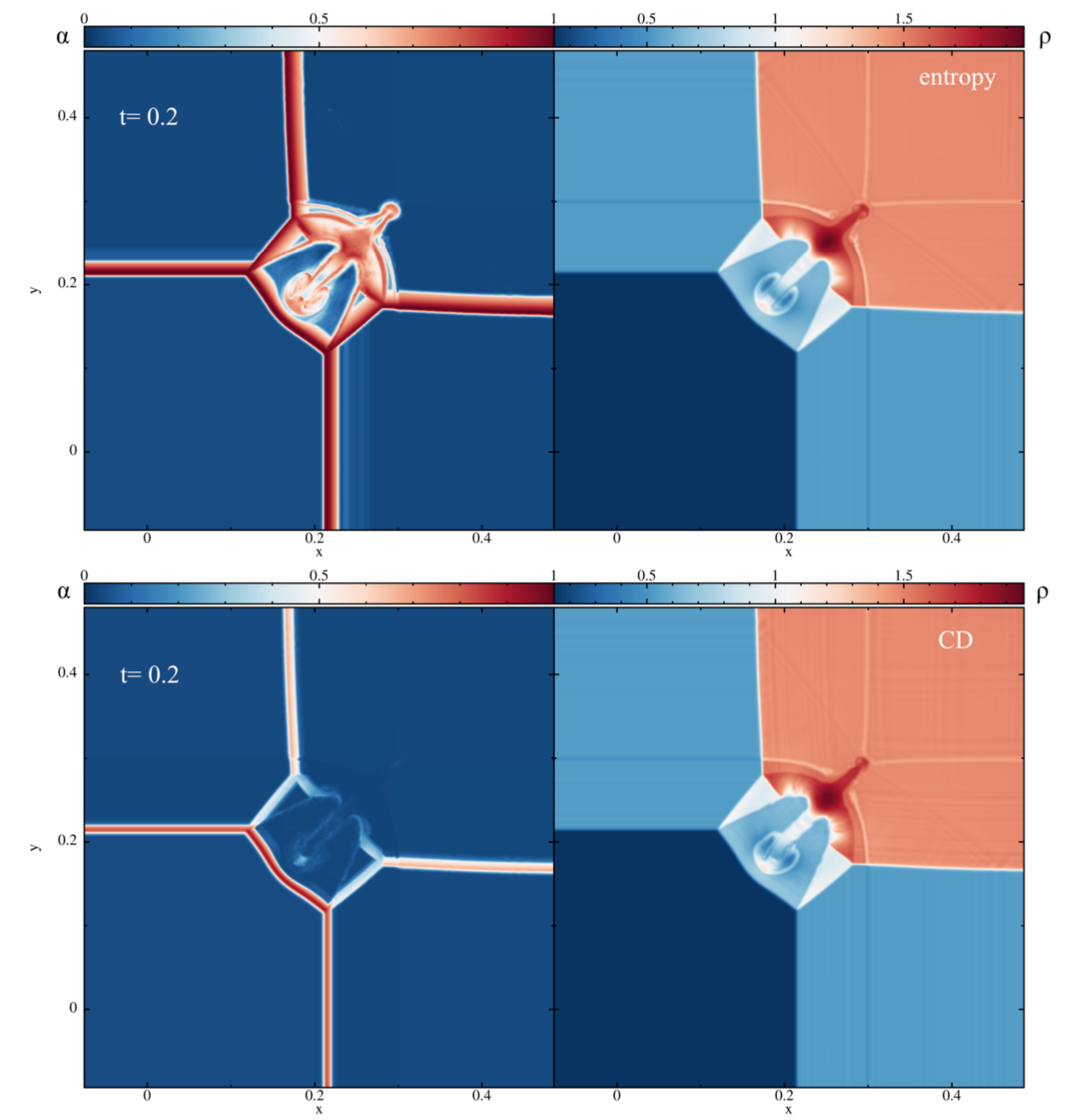}}
\vspace*{0cm}
\caption{Schulz-Rinne test 1 with the suggested entropy steering (top row) and the \CD-trigger
(bottom row, "CD"). In each row the dissipation parameter $\alpha$ is shown on the left and  the density $\rho$ on the right. 
Note the emerging asymmetries in the density in the bottom row since too little dissipation is triggered.}
\label{fig:SchulzRinne1_combined}
\end{figure*}
%-------------------------------------------------------------------------------
%
%-------------------------------------------------------------------------------
%%%% in original labelling this is **Test 12** %%%%%%%%%%%%
\begin{figure*}
\centerline{\includegraphics[width=2\columnwidth,angle=0]{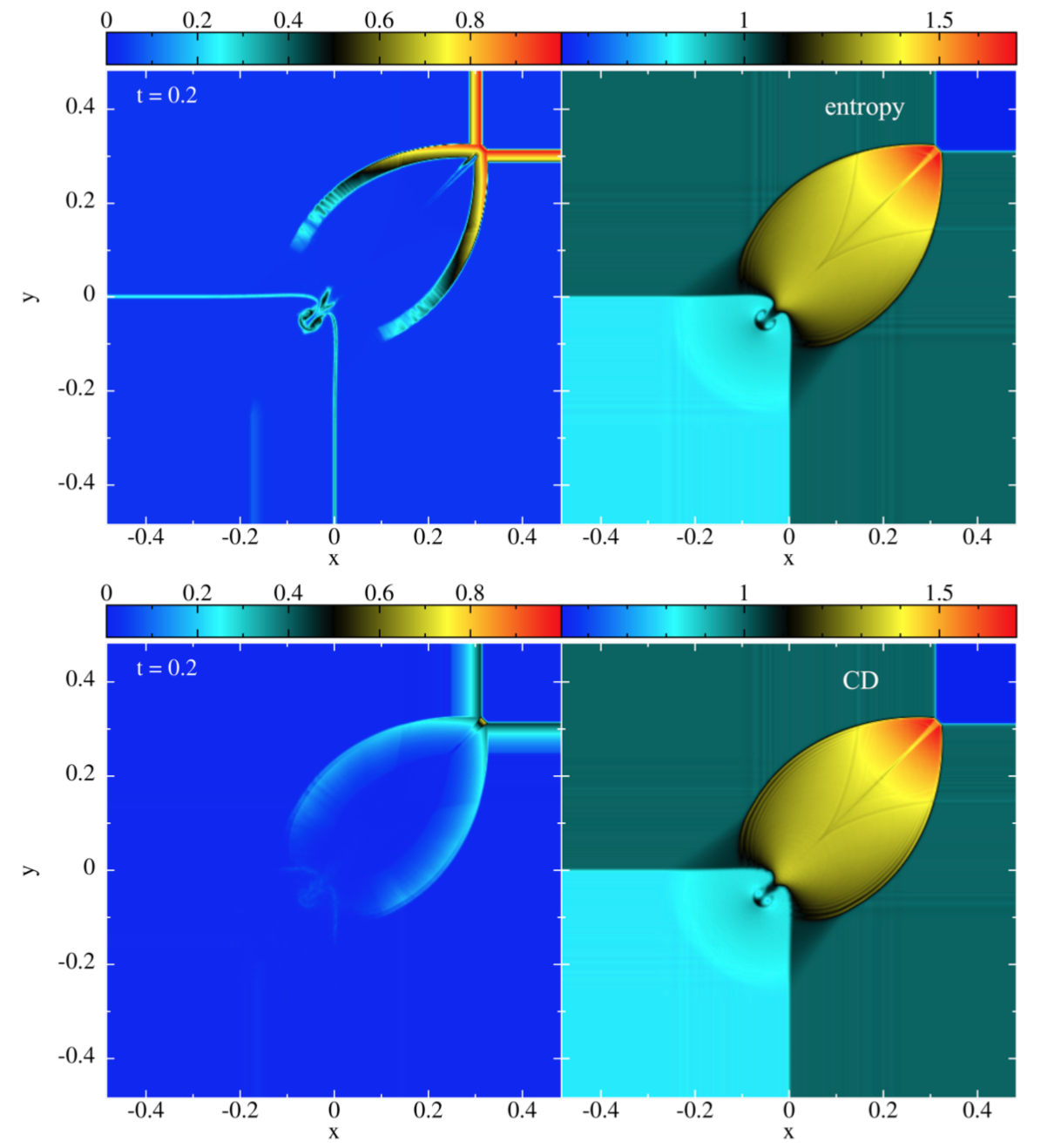}}
\vspace*{0cm}
\caption{Schulz-Rinne test 2 with the suggested entropy steering (top row) and the scheme suggested by Cullen \& Dehnen (2010)
(bottom row, "CD"). In each row the dissipation parameter $\alpha$ is shown on the right and density $\rho$  on the left.}
\label{fig:SchulzRinne2_combined}
\end{figure*}
%-------------------------------------------------------------

\subsection{Circular Blast}
As another benchmark we use a  three-dimensional shock-tube problem
suggested by \cite{toro99}. Similar to the Sedov test, we set up $200^3$ particles 
in the computational domain  $[-1,1]^3$. We first place them according to a centroidal Voronoi tesselation 
\citep{du99} and then perform 500 regularization sweeps. The fluid properties are assigned
according to
\be
(\rho,\vec{v},P)=
 \left\{
\begin{array}{l}
(1.000,0,0,0,1.0) \quad {\rm for \; \; r< 0.5}\\
(0.125,0,0,0,0.1) \quad {\rm else.}
\end{array}
\right.
\ee
The solution exhibits a spherical shock wave, a spherical contact surface traveling in the same direction
and a spherical rarefaction wave traveling towards the origin. Our solution ($xy$-plane) at
$t=0.2$ is shown in Fig.~\ref{fig:Riemann1} with  dissipation parameter $\alpha$ (left), density $\rho$, 
velocity $v$ and pressure $P$. While the "glass-like" initial particle distribution leaves some weak
imprint on the $\alpha$-values (left panel) all the physical quantities (panels 2-4) are essentially perfectly
spherically symmetric. The comparison of the \ma-result ($|y|<0.018, |z|< 0.018$) with the reference 
solution obtained with the Eulerian weighted average flux method with $400^3$ grid cells  \citep{toro99}
shows a very good agreement between both, see Fig.~\ref{fig:Riemann1_ref}, upper row. In the lower
row of this figure we show the \CD-trigger results. Again the agreement is very good, but since less dissipation is 
triggered there are larger velocity overshoots at the shock front.

\subsection{Schulz-Rinne  tests}
\begin{table}
\caption{Initial data for the Schulz-Rinne-type 2D Riemann problems}
\begin{tabular}{ l c | c | c | c | c | c | c |}
\hline
 & & SR1; contact point: $(0.3,0.3)$ & &\\
  \hline	
  \hline		
  variable & NW & NE & SW & SE \\ \hline
  $\rho$ &  0.5323 &  1.5000  & 0.1380 &  0.5323 \\
  $v_x$ &  1.2060 &  0.0000   & 1.2060 & 0.0000  \\
  $v_y$ &   0.0000 & 0.0000   & 1.2060 & 1.2060 \\
  $P$    &   0.3000 & 1.5000   & 0.0290 & 0.3000    \\
  \hline  
  & & SR2; contact point: $(0.0,0.0)$ & &\\
  \hline	
  variable & NW & NE & SW & SE \\ \hline
  $\rho$ &  1.0000 &  0.5313 & 0.8000  &  1.000 \\
  $v_x$ &   0.7276 &  0.0000 & 0.0000 & 0.0000  \\
  $v_y$ &   0.0000 & 0.0000  & 0.0000  &  0.7262\\
  $P$    &   1.0000  & 0.4000  & 1.0000 & 1.0000    \\
  \hline  
  
\end{tabular}
\label{tab:SR}
\end{table}
\cite{schulzrinne93a} designed a set of  challenging 2D Riemann problems in which 
four constant states meet at one corner. The initial conditions are chosen so that one 
elementary wave, either a shock,
a rarefaction or a contact discontinuity emerges from each interface and the subsequent evolution
leads to geometrically complex solutions. No exact solutions are known, but the benchmark tests
are often used and the results can be compared to other numerical solutions
\citep{schulzrinne93a,lax98,kurganov02,liska03}.
Here we show the results for two such tests, the initial conditions of which are 
given in Tab.~\ref{tab:SR}. Further tests of this type are shown in the \Ma code 
paper \citep{rosswog20a}\footnote{The  tests shown here are test 3 and 12 in the numbering
of \cite{kurganov02}.}.\\
These tests are rarely shown for SPH codes, in fact, we are only aware of the work 
by \cite{puri14} who show results for one such shock test
in a study of Godunov-SPH with different approximate Riemann solvers. Most of their implementations,
however, show serious artefacts in this test.
Since our code is intrinsically 3D, we simulate a slice thick enough so that the midplane is
unaffected by edge effects (we use here 10 particle layers in z-direction).  We
use 660 x 660  close-packed particles in the $xy$-plane between  $[x_c-0.5,x_c+0.5] \times [y_c-0.5,y_c+0.5]$, 
$(x_c,y_c)$ being the contact point of the quadrants, and we use
a polytropic exponent $\Gamma=1.4$.\\
 Fig.~\ref{fig:SchulzRinne1_combined} shows the results for test SR1 with the upper two panels showing 
 dissipation parameter $\alpha$ (left) and density (right) for the entropy trigger. The corresponding quantities
 for the \CD-trigger are shown in the lower two panels. The general features of the solution are captured 
 in both cases, but the \CD-trigger provides substantially less dissipation in the central vortex-like 
region which results in a noticeable lack of symmetry (lower right panel).\\
The results for the SR2 test are shown in Fig.~\ref{fig:SchulzRinne2_combined}. Again, the results for the
entropy trigger are given in the upper panels, those for the \CD-trigger in the lower ones.
The entropy scheme delivers sharp and noise-free density structures with
a high degree of symmetry and that are in very good agreement with the results from Eulerian 
approaches \citep{schulzrinne93a,lax98,kurganov02,liska03}. The CD scheme captures the 
overall features, but again triggers less dissipation which leads to the central vortex structure
being substantially less developed.
%-------------------------------------------------------------
\begin{figure*}
\hspace*{-0.cm}\includegraphics[width=2.2\columnwidth,angle=0]{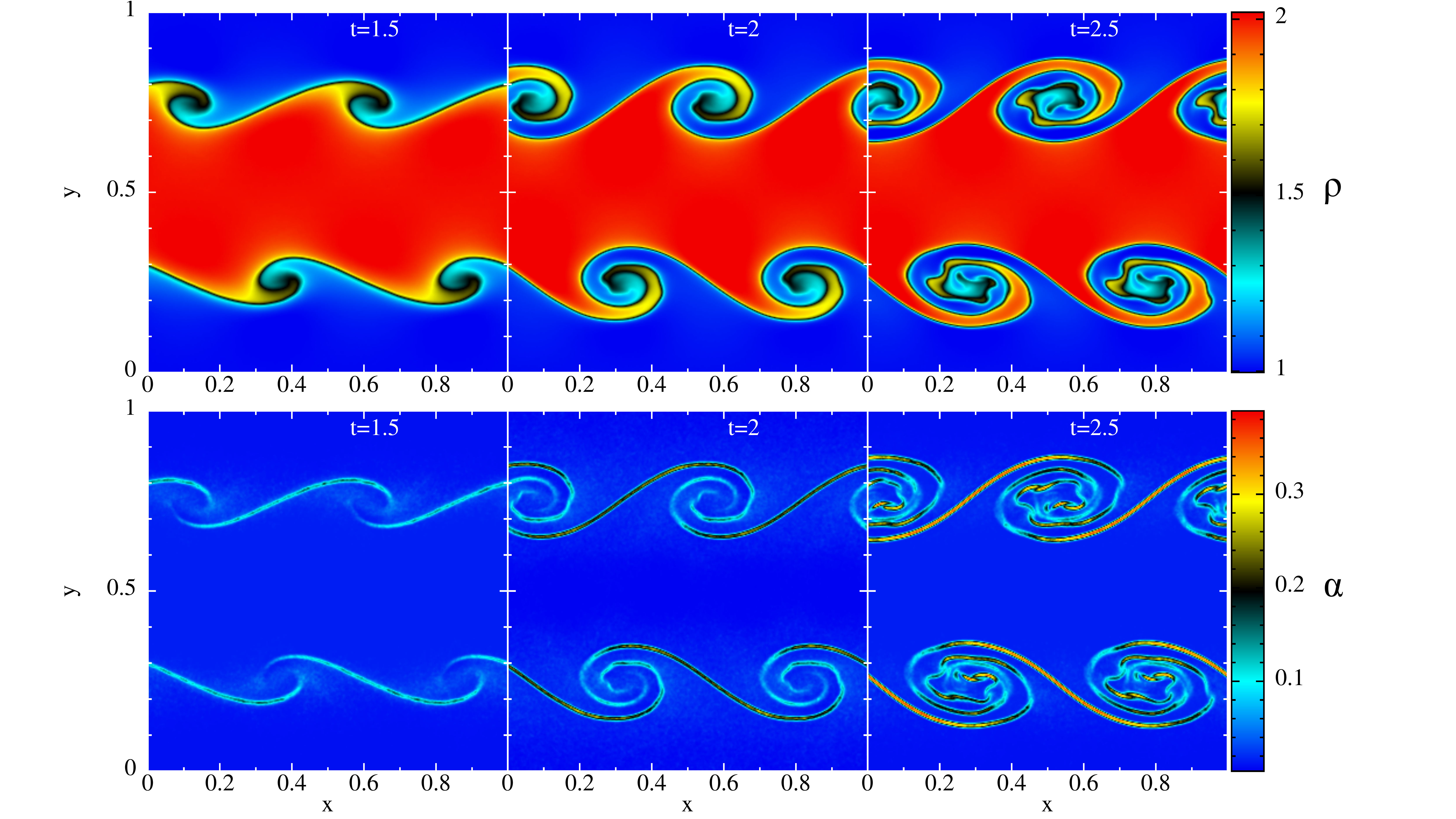}
\vspace*{0cm}
\caption{Kelvin-Helmholtz test, density $\rho$ is shown in top, dissipation parameter $\alpha$ in the bottom row.}
\label{fig:KH}
\end{figure*}
%-----------------------------------------------------------------------------
%
%-----------------------------------------------------------------------------
\begin{figure}
\hspace*{-0.1cm}\includegraphics[width=1.2\columnwidth,angle=0]{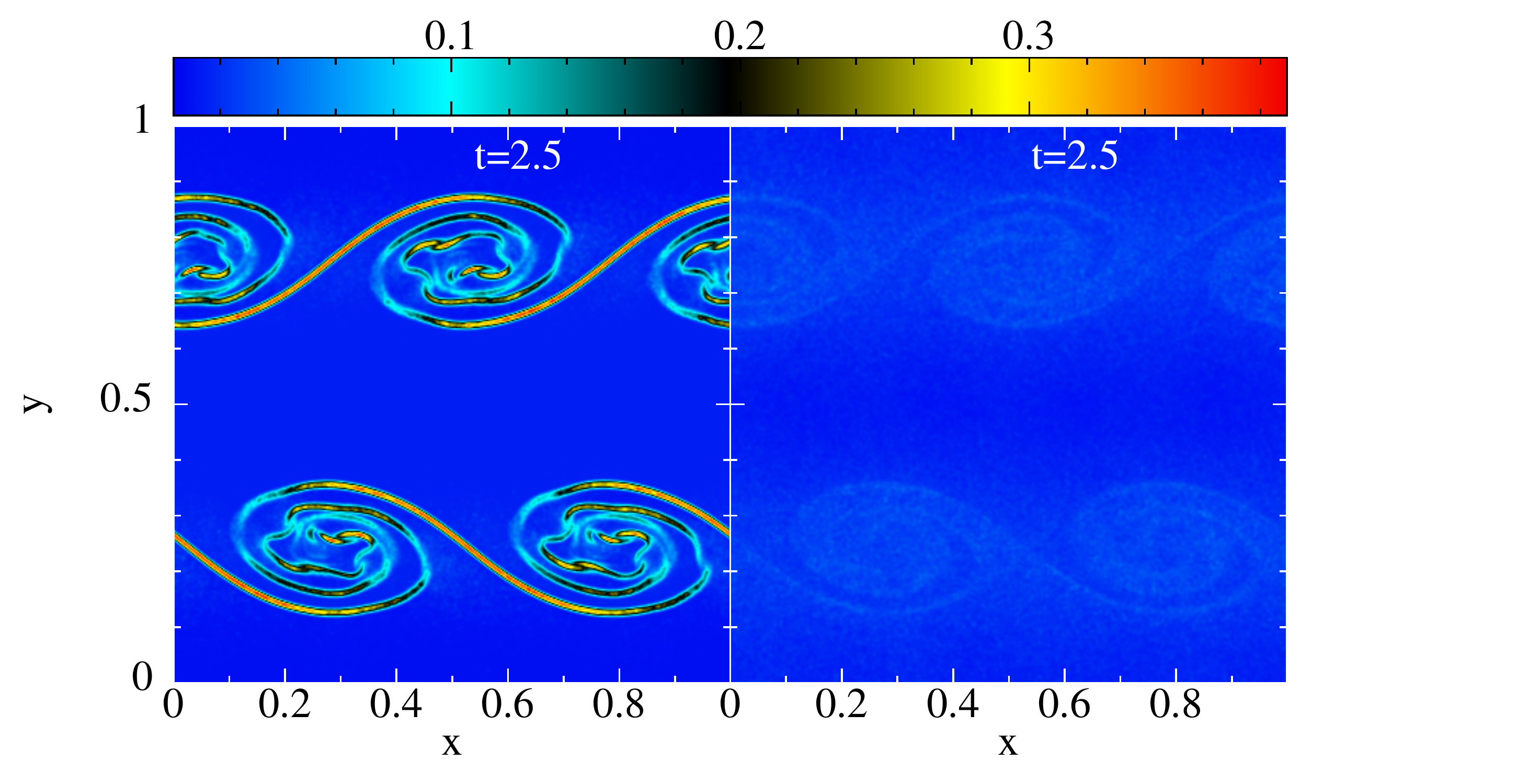}
\vspace*{0cm}
\caption{Comparison of the dissipation values $\alpha$ triggered by the entropy (left) and the \CD-method (right).}
\label{fig:KH_AV}
\end{figure}
%-------------------------------------------------------------------------------
%
%-----------------------------------------------------------------------------
\begin{figure}
\hspace*{-0.cm}\includegraphics[width=1.1\columnwidth,angle=0]{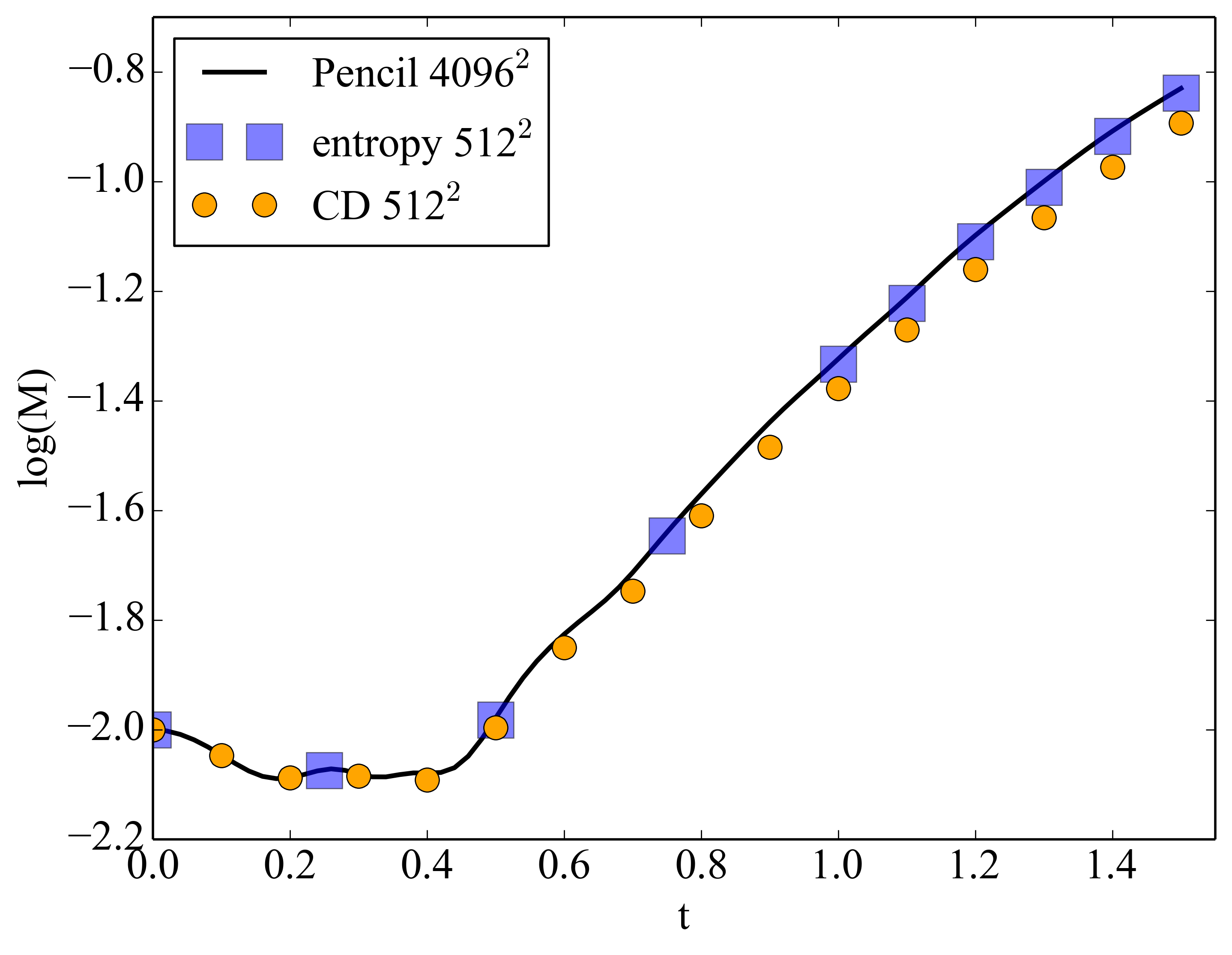}
\vspace*{0cm}
\caption{Kelvin-Helmholtz rate calculated as in McNally et al. (2012). The growth rate with the suggested entropy trigger is
shown as blue squares ("entropy"), with the \CD-trigger as orange ("CD"). As reference solution we take
a high resolution ($4096^2$ grid cells) simulation from the Pencil code.}
\label{fig:KH_growth}
\end{figure}
%-------------------------------------------------------------------------------

%-----------------------------------------------------------------------------
\begin{figure*}
\hspace*{-0.1cm}\includegraphics[width=14cm,angle=0]{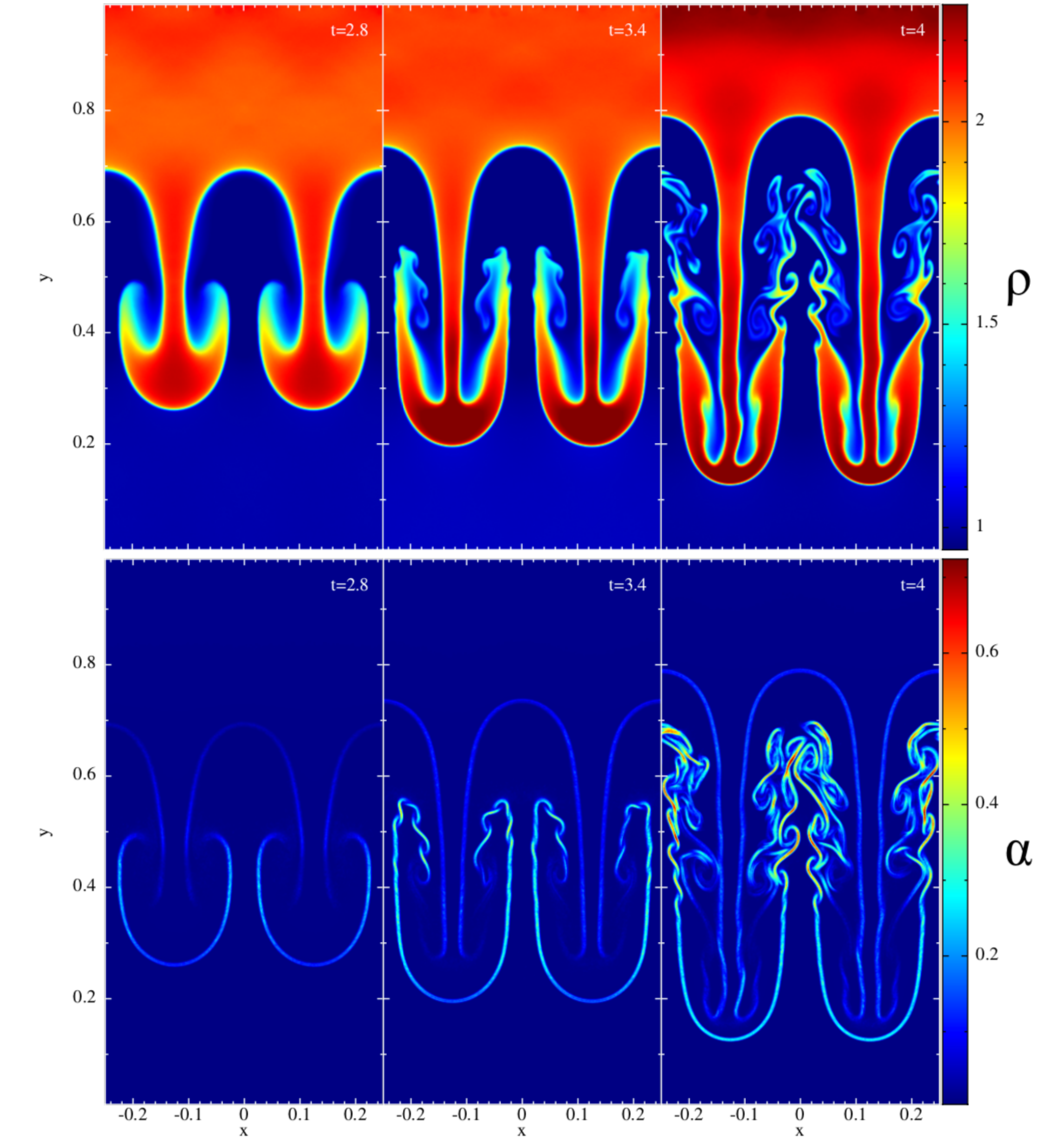}
\vspace*{0cm}
\caption{Rayleigh-Taylor test ($512 \times 1024$ particles). The top row shows the density $\rho$ and 
bottom row the evolution of the dissipation parameter $\alpha$ triggered by the suggested entropy criterion.}
\label{fig:RT}
\end{figure*}
%-------------------------------------------------------------------------------
%
%-----------------------------------------------------------------------------
\begin{figure}
\centerline{\includegraphics[width=1.2\columnwidth,angle=0]{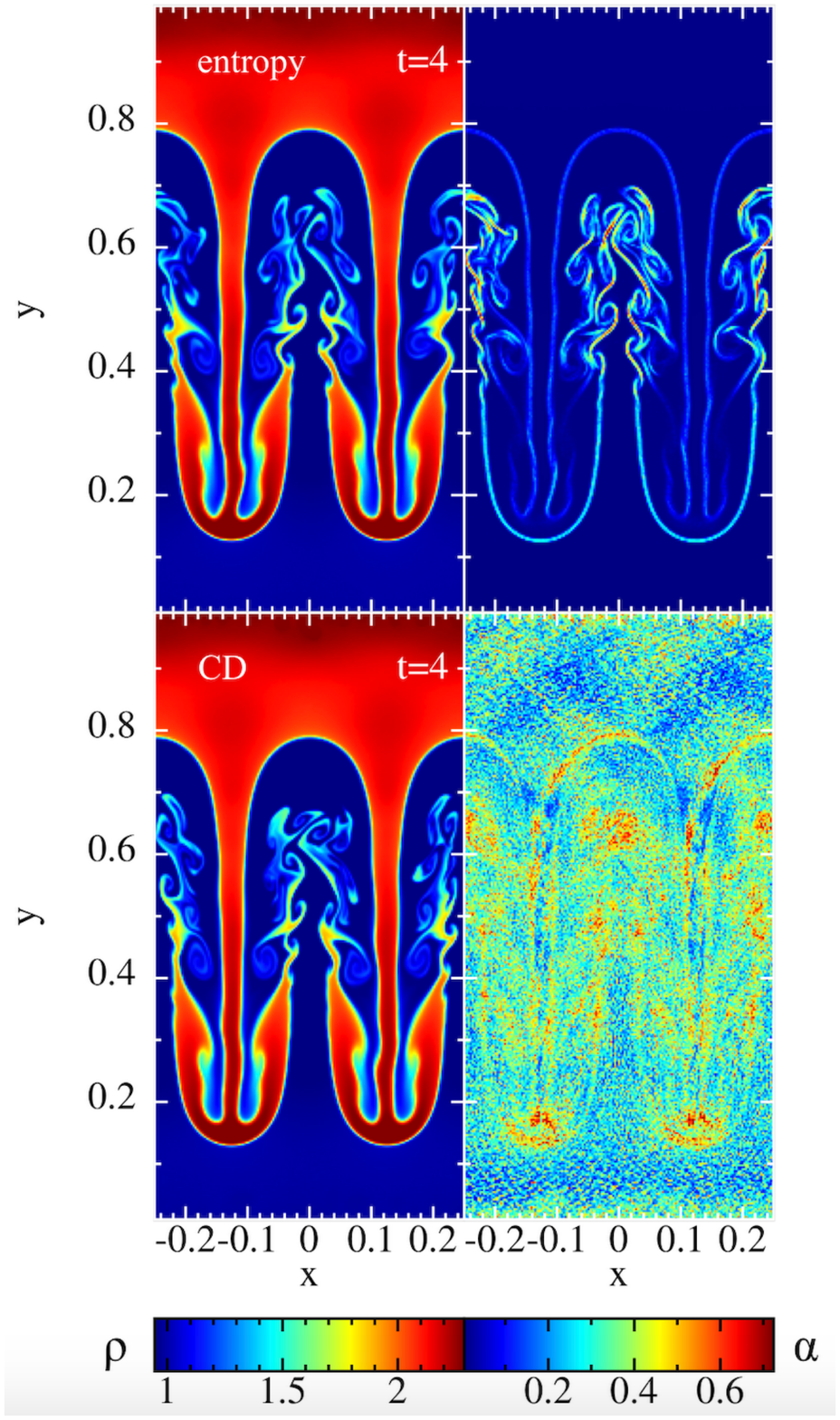}}
\vspace*{0cm}
\caption{Comparison of the triggered dissipation parameter values, entropy trigger (top) and \CD-trigger (bottom).}
\label{fig:RaTa_AV_en_CD}
\end{figure}
%-------------------------------------------------------------------------------
%
%-----------------------------------------------------------------------------
\begin{figure}
\centerline{\includegraphics[width=\columnwidth,angle=0]{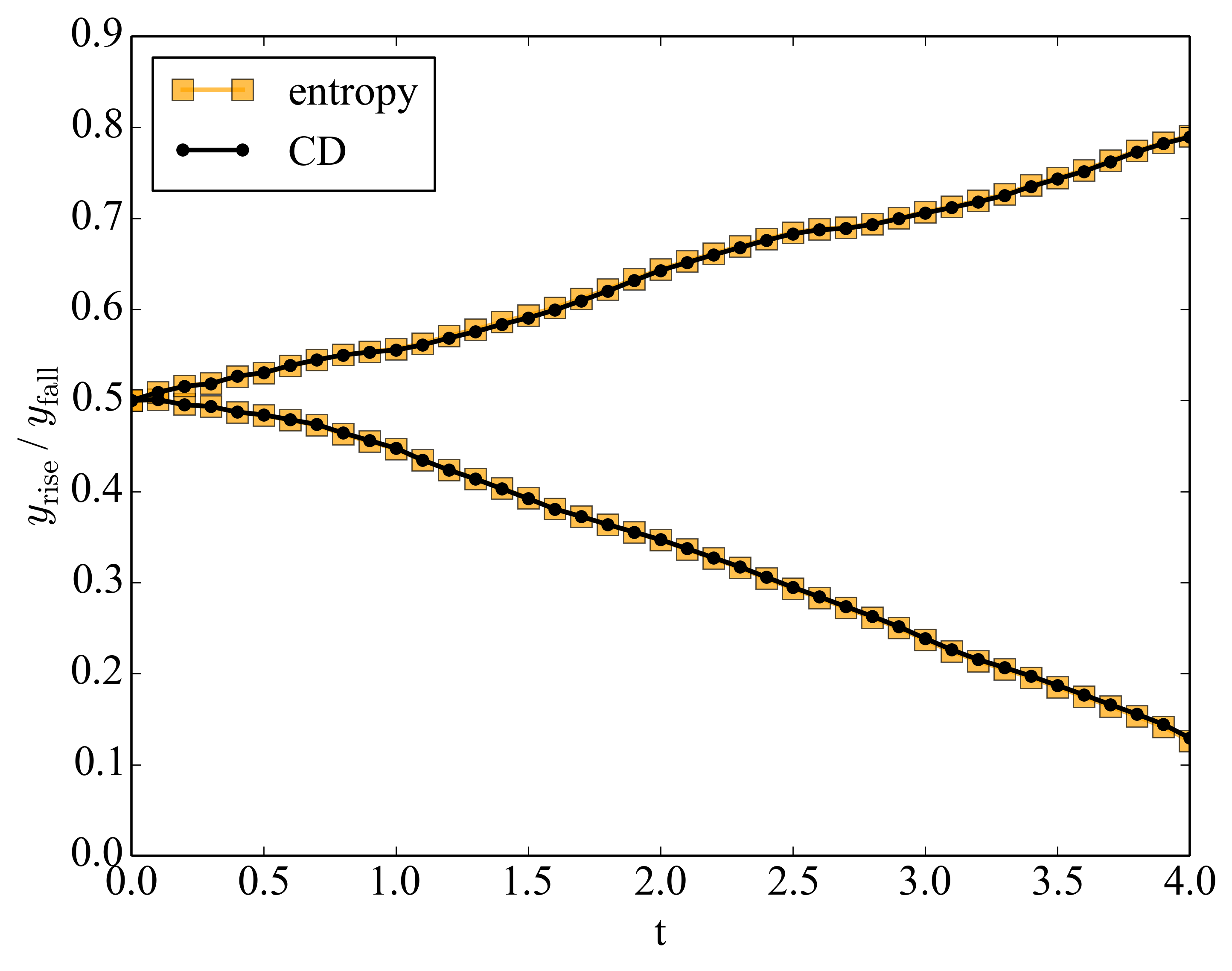}}
\vspace*{0cm}
\caption{Measure of the $y-$ positions of the density interface in the rising (at $x=0$, $y_{\rm rise}$)
and down-sinking part of the flow (at $x=0.125$, $y_{\rm fall}$) for both
the entropy- (orange) and the \CD-steering (black).}
\label{fig:y_rise_fall}
\end{figure}
%-------------------------------------------------------------------------------

\subsection{Kelvin Helmholtz}
An interesting question is how much dissipation is triggered 
in an overall smooth test such as a Kelvin-Helmholtz instability. To find out we set up
a test  similar to \cite{mcnally12} and \cite{frontiere17}.  The test is initialized as:
\be
\rho(y)=
 \left\{
\begin{array}{l}
\rho_1 - \rho_m e^{(y - 0.25)/\Delta} \quad {\rm for \; \; 0.00 < y < 0.25}\\
\rho_2 + \rho_m e^{(0.25 - y)/\Delta} \quad {\rm for \; \; 0.25 \le y < 0.50}\\
\rho_2 + \rho_m e^{(y - 0.75)/\Delta} \quad {\rm for \; \; 0.50 \le y < 0.75}\\
\rho_1 - \rho_m e^{(0.75 - y)/\Delta} \quad {\rm for \; \; 0.75 \le y < 1.00}\\
\end{array}
\right.
\ee
where $\rho_1= 1$, $\rho_2= 2$, $\rho_m= (\rho_1 - \rho_2)/2$ and $\Delta= 0.025$
and the initial velocities are
\be
v_x(y)=
 \left\{
\begin{array}{l}
v_1 - v_m e^{(y - 0.25)/\Delta} \quad {\rm for \; \; 0.00 < y < 0.25}\\
v_2 + v_m e^{(0.25 - y)/\Delta} \quad {\rm for \; \; 0.25 \le y < 0.50}\\
v_2 + v_m e^{(y - 0.75)/\Delta} \quad {\rm for \; \; 0.50 \le y < 0.75}\\
v_1 - v_m e^{(0.75 - y)/\Delta} \quad {\rm for \; \; 0.75 \le y < 1.00}\\
\end{array}
\right.
\ee
with $v_1$= 0.5, $v_2= -0.5$, $v_m= (v_1-v_2)/2$. A small velocity perturbation in
$y$-direction is introduced as $v_y=  0.01 \sin(2\pi x/\lambda)$ with the perturbation
wave length  $\lambda= 0.5$. The test is performed with a polytropic equation of state 
with exponent $\Gamma=5/3$. The test is set up in quasi-2D with 20 slices of 512 x 512 
particles which, for simplicity, are arranged in a simple cubic lattice. Here we focus 
exclusively on the topic of this
paper, the dissipation trigger, for more details of the setup and the analysis of the
\Ma performance we refer to \cite{rosswog20a}. The result at $t=1.5, 2.0$ and
$2.5$ is shown in Fig.~\ref{fig:KH}. Our dissipation scheme triggers a floor value
of $\alpha\approx 0.015$. In the shear interfaces where
the particle lattices shear along one another sharply localized lines with values of up to $\alpha\approx0.4$
are triggered. The average dissipation parameter value at $t=2.5$ is $\bar{\alpha}= 0.09$, i.e.
the $\alpha-$values are substantially lower than the standard value of unity. \\
We have repeated this test with the \cd-trigger. The density evolution is visually
very similar to the one shown in Fig.~\ref{fig:KH}, the only noticeable difference is that
the Kelvin-Helmholtz billows at late times are less symmetric than for the entropy trigger. 
The $\alpha$ values (at t= 2.5) are shown in Fig.~\ref{fig:KH_AV}:
the \CD method hardly switches on anywhere and reaches even in the Kelvin-Helmholtz billows
only values of $\alpha\sim0.03$.
We have also measured the growth rate of the instability, calculated exactly as in 
\cite{mcnally12}.  As a reference solution we use a high resolution calculation ($4096^2$ cells) 
obtained by the PENCIL code \citep{brandenburg02}. Both our cases show a healthy growth rate, but
the entropy method is noticeably closer to the reference solution, likely because noise is more efficiently
suppressed.

\subsection{Rayleigh-Taylor Instability}
As a last example we show the results for a commonly used Rayleigh-Taylor test 
\citep{abel11,hopkins15a,frontiere17}. Again, the focus is on the amount of dissipation that is triggered,
more details on this test can be found in original code paper \citep{rosswog20a}. We adopt a quasi-2D setup using the full 
3D code in a $xy$-domain of $[-0.25,0.25] \times [0,1]$ and use 10 layers of particles in the z-direction. 
The initial density is set up as
\be
\rho(y)= \rho_b + \frac{\rho_t-\rho_b}{1+\exp[-(y-y_t)/\Delta]}
\ee
with $\rho_t=2$, $\rho_b=1$, transition width $\Delta=0.025$ and transition coordinate $y_t=0.5$. 
The interface is perturbed as
\be
v_y(x,y)= \delta v_{y,0} [1 + \cos(8\pi x)][1 + \cos(5\pi(y-y_t))]
\ee
for $y$ in $[0.3,0.7]$ with an initial amplitude $\delta v_{y,0}=0.025$. The equilibrium pressure profile
is given by
\be
P (y) = P_0 -  g \rho(y) [y - y_t],
\ee
where $P_0= \rho_t/\Gamma$ and polytropic exponent is chosen as $\Gamma=1.4$. A constant acceleration $\vec{g}= -0.5 \hat{e}_y$
is applied. We show snapshots at $t= 2.8, 3.4$ and 4.0 in Fig.~\ref{fig:RT} with density in the upper row
and the corresponding dissipation parameters in the lower row. Throughout most of the computational domain the value of
$\alpha$ is very low ($\sim 0.01$), only in the sharp edges of the rising plumes values up to $\approx 0.7$ are reached. 
Overall, our results  in this test are very similar to those obtained with Lagrangian finite volume particle methods \citep{hopkins15a} 
and SPH-methods  \citep{frontiere17} based on the reproducing kernel methodology \citep{liu95}.\\
We have repeated this test with the  \CD-trigger. We find that it actually triggers more dissipation 
in this test than the new approach, see Fig.~\ref{fig:RaTa_AV_en_CD}, right panels, but the effects 
are benign since the evolution (at least with our approach) is not very
sensitive to the exact values of $\alpha$. Visually, the density evolution (left panels) is very similar between the
two approaches. We further compare, more quantitatively, the position of the fluid interface
in the central rising and the down-sinking parts of the flow. Practically, we use 
bisections  to track the density value $(\rho_t+\rho_b)/2= 1.5$ along the y-axis, $y_{\rm rise}$, 
and along $x= 0.125$ (the right down-sinking "mushroom"), $y_{\rm fall}$, to capture the interface positions. 
These interface positions are shown in Fig.~\ref{fig:y_rise_fall}, but, as expected from Fig.~\ref{fig:RaTa_AV_en_CD}, there are 
hardly any differences noticeable.

\section{Summary}
\label{sec:summary}
In this paper we have explored a novel way to steer dissipation in SPH simulations.
Rather than triggering on the velocity divergence \citep{morris97} or its
time derivative \citep{cullen10}, we trigger on local violations of exact
entropy conservation. Such violations can be caused by particles entering 
a shock front or by numerical noise. We find the additional triggering on
noise very beneficial in calming down post-shock regions and in
resolving complex fluid structures as they emerge e.g. 
in Schulz-Rinne test cases. Triggers on noise, in addition to a shock trigger, 
had been suggested in earlier work \citep{rosswog15b}, but the new scheme 
discussed here is much simpler and the same triggering mechanism takes care of
both shocks and noise.
The new scheme switches on robustly in shocks, moderately and only very locally in the
case of noise and it triggers hardly any dissipation  ($\alpha \sim 0.01$ for our parameter choice) in  
smooth regions of the flow. The suggested method is very robust, trivial to implement in existing SPH codes 
and does not require any noticeable computational effort. 

\section*{Acknowledgements}
Some of the figures of this article were 
produced with the visualization software SPLASH \citep{price07a}. 
This work has been supported by the Swedish Research 
Council (VR) under grant number 2016- 03657\_3, by 
the Swedish National Space Board under grant number 
Dnr. 107/16 and by the research environment grant "Gravitational Radiation and Electromagnetic Astrophysical
Transients (GREAT)" funded by the Swedish Research 
council (VR) under Dnr 2016-06012. We gratefully 
acknowledge support from COST Action CA16104 
"Gravitational waves, black holes and fundamental physics" (GWverse) and from COST Action CA16214
"The multi-messenger physics and astrophysics of neutron stars" (PHAROS).
The simulations for this paper were performed on the facilities of the North-German Supercomputing Alliance (HLRN),
and on the SNIC resources Tetralith and Beskow.

\bibliographystyle{mn2e}
\bibliography{astro_SKR.bib}
\end{document}